\documentstyle[amssymb,floats,aps,prb,epsf]{revtex}
\epsfclipon
\begin{document}
\twocolumn[\hsize\textwidth\columnwidth\hsize\csname
@twocolumnfalse\endcsname

\draft

\title{Enhancement of superconducting transition temperature by
the additional second neighbor hopping $t'$ in the $t$-$J$ model}
\author{Shiping Feng and Tianxing Ma}
\address{Department of Physics, Beijing Normal University, Beijing
100875, China}
\maketitle
\begin{abstract}
Within the kinetic energy driven superconducting mechanism, the
effect of the additional second neighbor hopping $t'$ on the
superconducting state of the $t$-$J$ model is discussed. It is
shown that $t'$ plays an important role in enhancing the
superconducting transition temperature of the $t$-$J$ model. It is
also shown that the superconducting-state of cuprate
superconductors is the conventional Bardeen-Cooper-Schrieffer
like, so that the basic Bardeen-Cooper-Schrieffer formalism is
still valid in quantitatively reproducing the doping dependence of
the superconducting gap parameter and superconducting transition
temperature, and electron spectral function at ($\pi$,0) point,
although the pairing mechanism is driven by the kinetic energy by
exchanging dressed spin excitations.
\end{abstract}


]

\bigskip

\narrowtext

After intensive investigations over more than a decade, it has now
become clear that although the physical properties of cuprate
superconductors in the normal-state are fundamentally different
from these of the conventional metals \cite{kastner}, the
superconducting (SC)-state of cuprate superconductors is still
associated with the formation of the electron Cooper pairs
\cite{tsuei} as in the conventional superconductors. In the
conventional metals, superconductivity results when electrons pair
up into Cooper pairs, which is mediated by the interaction of
electrons with phonons \cite{bcs}. As a result, the pairing in the
conventional superconductors is always related with an increase in
kinetic energy which is overcompensated by the lowering of
potential energy \cite{chester}. However, it has been argued that
the form of the electron Cooper pairs is determined by the need to
reduce the frustrated kinetic energy in doped cuprates
\cite{anderson2,molegraaf}, i.e., the strong frustration of the
kinetic energy in the normal-state is partially relieved upon
entering the SC-state. By virtue of systematic studies using the
nuclear magnetic resonance, and muon spin rotation techniques,
particularly the inelastic neutron scattering, it has been well
established that the antiferromagnetic (AF) short-range
correlation (AFSRC) coexists with the SC-state in the whole SC
regime \cite{yamada,wakimoto}, which provide a clear link between
the SC pairing mechanism and magnetic excitations. Moreover, it
has been shown \cite{ding2} that although the SC pairing mechanism
of cuprate superconductors is beyond the conventional
electron-phonon mechanism, the SC-state is the conventional
Bardeen-Cooper-Schrieffer (BCS) like \cite{bcs}, so that the basic
BCS formalism is still valid in discussions of the electron
spectral properties \cite{ding2}.

Very soon after the discovery of superconductivity in doped
cuprates, Anderson \cite{anderson1} suggested that the essential
physics of doped cuprates is contained in the $t$-$J$ model on a
square lattice. This followed from the experiments that cuprate
superconductors are doped antiferromagnets, where the common
features are the presence of the square lattice CuO$_{2}$ planes
\cite{kastner} and a similar phase diagram as a function of the
doping concentration \cite{tallon}. Since then much effort has
concentrated on the unusual normal-state and SC mechanism within
the $t$-$J$ model \cite{anderson2,laughlin}. Based on the
charge-spin separation (CSS) fermion-spin theory \cite{feng1}, we
\cite{feng2} have developed a kinetic energy driven SC mechanism
within the $t$-$J$ model. It is shown \cite{feng2} that the
dressed holons interact occurring directly through the kinetic
energy by exchanging the spin excitations, leading to a net
attractive force between the dressed holons, then the electron
Cooper pairs originating from the dressed holon pairing state are
due to the charge-spin recombination, and their condensation
reveals the SC ground-state. This SC-state is controlled by both
SC gap function and quasiparticle coherence, and the maximal SC
transition temperature occurs around the optimal doping, then
decreases in both underdoped and overdoped regimes \cite{feng3}.
However, the simple $t$-$J$ model can not be regarded as a
comprehensive model for the quantitative comparison with cuprate
superconductors. It has been shown \cite{well} from the angle
resolved photoemission spectroscopy (ARPES) experiments that
although the highest energy filled electron band is well described
by the $t$-$J$ model in the direction between the $[0,0]$ point
and the $[\pi,\pi]$ point in the momentum space, but both
experimental data near $[\pi,0]$ point and overall dispersion may
be properly accounted by generalizing the $t$-$J$ model to include
the second- and third-nearest neighbors hopping terms $t'$ and
$t''$. Moreover, the experimental analysis \cite{tanaka} shows
that the SC transition temperature for different families of
cuprate superconductors is strongly correlated with $t'$. In this
Letter, we discuss the effect of the additional second neighbor
hopping $t'$ on the SC-state of the $t$-$J$ model within the
framework of the kinetic energy driven SC mechanism \cite{feng2}.
Our result shows that the SC-state of cuprate superconductors is
the conventional BCS like \cite{bcs}, so that the basic BCS
formalism is still valid in quantitatively reproducing the doping
dependence of the effective SC gap parameter and SC transition
temperature, and electron spectral function at $[\pi,0]$ point,
although the pairing mechanism is driven by the kinetic energy by
exchanging dressed spin excitations, and other exotic magnetic
properties \cite{feng3} are beyond the BCS theory. Our result also
shows that the additional second neighbor hopping $t'$ plays an
important role in enhancing the SC transition temperature of the
$t$-$J$ model and in determining the correct position of the SC
quasiparticle peak of the electron spectral function at $[\pi,0]$
point.

We start from the $t$-$t'$-$J$ model on a square lattice
\cite{anderson1,well},
\begin{eqnarray}
H&=&-t\sum_{i\hat{\eta}\sigma}C^{\dagger}_{i\sigma}
C_{i+\hat{\eta}\sigma}+t'\sum_{i\hat{\tau}\sigma}
C^{\dagger}_{i\sigma}C_{i+\hat{\tau}\sigma}+\mu\sum_{i\sigma}
C^{\dagger}_{i\sigma}C_{i\sigma}\nonumber \\
&+&J\sum_{i\hat{\eta}}{\bf S}_{i} \cdot {\bf S}_{i+\hat{\eta}},
\end{eqnarray}
supplemented by the local constraint $\sum_{\sigma}
C^{\dagger}_{i\sigma} C_{i\sigma}\leq 1$ to avoid the double
occupancy, where $\hat{\eta}=\pm\hat{x},\pm \hat{y}$, $\hat{\tau}=
\pm\hat{x}\pm\hat{y}$, $C^{\dagger}_{i\sigma}$ ($C_{i\sigma}$) is
the electron creation (annihilation) operator, ${\bf S}_{i}=
C^{\dagger}_{i}{\vec\sigma} C_{i}/2$ is spin operator with
${\vec\sigma}=(\sigma_{x}, \sigma_{y},\sigma_{z})$ as Pauli
matrices, and $\mu$ is the chemical potential. The strong electron
correlation in the $t$-$t'$-$J$ model manifests itself by the
electron single occupancy local constraint \cite{anderson1}, which
can be treated properly in analytical calculations within the CSS
fermion-spin theory \cite{feng1}, where the constrained electron
operators are decoupled as $C_{i\uparrow}= h^{\dagger}_{i\uparrow}
S^{-}_{i}$ and $C_{i\downarrow}= h^{\dagger}_{i\downarrow}
S^{+}_{i}$, with the spinful fermion operator $h_{i\sigma}=
e^{-i\Phi_{i\sigma}}h_{i}$ describes the charge degree of freedom
together with some effects of the spin configuration
rearrangements due to the presence of the doped hole itself
(dressed holon), while the spin operator $S_{i}$ describes the
spin degree of freedom (dressed spin), then the electron local
constraint for the single occupancy, $\sum_{\sigma}
C^{\dagger}_{i\sigma}C_{i\sigma}=S^{+}_{i} h_{i\uparrow}
h^{\dagger}_{i\uparrow}S^{-}_{i}+ S^{-}_{i}h_{i\downarrow}
h^{\dagger}_{i\downarrow}S^{+}_{i}=h_{i} h^{\dagger}_{i}(S^{+}_{i}
S^{-}_{i}+S^{-}_{i}S^{+}_{i})=1- h^{\dagger}_{i}h_{i}\leq 1$, is
satisfied in analytical calculations. It has been shown that these
dressed holon and spin are gauge invariant \cite{feng1}, and in
this sense, they are real and can be interpreted as the physical
excitations \cite{laughlin}. Although in common sense
$h_{i\sigma}$ is not a real spinful fermion, it behaves like a
spinful fermion. In this CSS fermion-spin representation, the
low-energy behavior of the $t$-$t'$-$J$ model (1) can be expressed
as,
\begin{eqnarray}
H&=&-t\sum_{i\hat{\eta}}(h_{i\uparrow}S^{+}_{i}
h^{\dagger}_{i+\hat{\eta}\uparrow}S^{-}_{i+\hat{\eta}}+
h_{i\downarrow}S^{-}_{i}h^{\dagger}_{i+\hat{\eta}\downarrow}
S^{+}_{i+\hat{\eta}})\nonumber \\
&+&t'\sum_{i\hat{\tau}}(h_{i\uparrow}S^{+}_{i}
h^{\dagger}_{i+\hat{\tau}\uparrow}S^{-}_{i+\hat{\tau}}+
h_{i\downarrow}S^{-}_{i}h^{\dagger}_{i+\hat{\tau}\downarrow}
S^{+}_{i+\hat{\tau}}) \nonumber \\
&-&\mu\sum_{i\sigma}h^{\dagger}_{i\sigma} h_{i\sigma}+J_{{\rm
eff}}\sum_{i\hat{\eta}}{\bf S}_{i}\cdot {\bf S}_{i+\hat{\eta}},
\end{eqnarray}
with $J_{{\rm eff}}=(1-x)^{2}J$, and $x=\langle
h^{\dagger}_{i\sigma}h_{i\sigma}\rangle=\langle h^{\dagger}_{i}
h_{i}\rangle$ is the hole doping concentration. As a consequence,
the kinetic energy terms in the $t$-$t'$-$J$ model have been
expressed as the dressed holon-spin interactions, which reflects
that even the kinetic energy terms in the $t$-$t'$-$J$ Hamiltonian
have strong Coulombic contributions due to the restriction of no
doubly occupancy of a given site, and therefore dominate the
essential physics of doped cuprates.

ARPES measurements \cite{shen1} show that in the real space the
gap function and pairing force have a range of one lattice
spacing, which indicates that the order parameter for the electron
Cooper pair can be expressed as,
\begin{eqnarray}
\Delta &=&\langle
C^{\dagger}_{i\uparrow}C^{\dagger}_{i+\hat{\eta}\downarrow}-
C^{\dagger}_{i\downarrow}C^{\dagger}_{i+\hat{\eta}\uparrow}\rangle
\nonumber \\
&=&\langle h_{i\uparrow}h_{i+\hat{\eta}\downarrow}S^{+}_{i}
S^{-}_{i+\hat{\eta}}-h_{i\downarrow}h_{i+\hat{\eta}\uparrow}
S^{-}_{i}S^{+}_{i+\hat{\eta}}\rangle .
\end{eqnarray}
In the doped regime without the AF long-range order (AFLRO), the
dressed spins form a disordered spin liquid state, where the
dressed spin correlation function $\langle S^{+}_{i}
S^{-}_{i+\hat{\eta}}\rangle=\langle S^{-}_{i}S^{+}_{i+\hat{\eta}}
\rangle$, then the order parameter for the electron Cooper pair in
Eq. (3) can be written as $\Delta=-\langle S^{+}_{i}
S^{-}_{i+\hat{\eta}}\rangle \Delta_{h}$, with the dressed holon
pairing order parameter $\Delta_{h}= \langle
h_{i+\hat{\eta}\downarrow}h_{i\uparrow}-h_{i+\hat{\eta}\uparrow}
h_{i\downarrow}\rangle$, which shows that the SC order parameter
of the electron Cooper pair is related to the dressed holon
pairing amplitude, and is proportional to the number of doped
holes, and not to the number of electrons. However, in the extreme
low doped regime with AFLRO, where the dressed spin correlation
function $\langle S^{+}_{i}S^{-}_{i+\hat{\eta}}\rangle\neq\langle
S^{-}_{i}S^{+}_{i+\hat{\eta}}\rangle$, then the conduct is
disrupted by AFLRO, and therefore there is no mixing of
superconductivity and AFLRO \cite{bozovic}. In the case without
AFLRO, we \cite{feng2,feng3} have shown within the Eliashberg's
strong coupling theory \cite{eliashberg} that the dressed
holon-spin interaction can induce the dressed holon pairing state
(then the electron Cooper pairing state) by exchanging dressed
spin excitations in the higher power of the hole doping
concentration. Following our previous discussions based on the
$t$-$J$ model \cite{feng2,feng3}, the self-consistent equations
that satisfied by the full dressed holon diagonal and off-diagonal
Green's functions in the present $t$-$t'$-$J$ model are obtained
as,
\begin{mathletters}
\begin{eqnarray}
g(k)&=&g^{(0)}(k)\nonumber \\
&+&g^{(0)}(k)[\Sigma^{(h)}_{1}(k)g(k)-
\Sigma^{(h)}_{2}(-k)\Im^{\dagger}(k)], \\
\Im^{\dagger}(k)&=&g^{(0)}(-k)[\Sigma^{(h)}_{1}(-k)
\Im^{\dagger}(-k)+\Sigma^{(h)}_{2}(-k)g(k)],
\end{eqnarray}
\end{mathletters}
respectively, where the four-vector notation $k=({\bf k},
i\omega_{n})$, and the mean-field (MF) dressed holon diagonal
Green's function \cite{feng1} $g^{(0)-1}(k)= i\omega_{n}-
\xi_{{\bf k}}$, with the MF dressed holon excitation spectrum
$\xi_{{\bf k}}=Zt\chi_{1}\gamma_{{\bf k}}-Zt'\chi_{2}\gamma_{{\bf
k}}'-\mu$, where $\gamma_{{\bf k}}=(1/Z)\sum_{\hat{\eta}}e^{i{\bf
k}\cdot \hat{\eta}}$, $\gamma_{{\bf k}}'=(1/Z)\sum_{\hat{\tau}}
e^{i{\bf k} \cdot\hat{\tau}}$, $Z$ is the number of the nearest
neighbor or second-nearest neighbor sites, the spin correlation
functions $\chi_{1}=\langle S_{i}^{+} S_{i+\hat{\eta}}^{-}\rangle$
and $\chi_{2}=\langle S_{i}^{+} S_{i+\hat{\tau}}^{-}\rangle$,
while the dressed holon self-energies are obtained from the
dressed spin bubble as,
\begin{mathletters}
\begin{eqnarray}
\Sigma^{(h)}_{1}(k)&=&{1\over N^{2}}\sum_{{\bf p,p'}} (Zt
\gamma_{{\bf p+p'+k}}-Zt'\gamma_{{\bf p+p'+k}}')^{2}\nonumber \\
&\times& {1\over \beta}\sum_{ip_{m}}g(p+k){1\over\beta}
\sum_{ip'_{m}} D^{(0)}(p')D^{(0)}(p'+p), \\
\Sigma^{(h)}_{2}(k)&=&{1\over N^{2}}\sum_{{\bf p,p'}} (Zt
\gamma_{{\bf p+p'+k}}-Zt'\gamma_{{\bf p+p'+k}}')^{2}\nonumber \\
&\times& {1\over \beta} \sum_{ip_{m}}\Im (-p-k){1\over\beta}
\sum_{ip'_{m}}D^{(0)}(p') D^{(0)}(p'+p),
\end{eqnarray}
\end{mathletters} where $p=({\bf p},ip_{m})$, $p'=({\bf p'},
ip_{m}')$, $N$ is the number of sites, and the MF dressed spin
Green's function \cite{feng1}, $D^{(0)-1}(p)=[(ip_{m})^{2}-
\omega_{{\bf p} }^{2}]/B_{{\bf p}}$, with $B_{{\bf p}}=2
\lambda_{1}(A_{1} \gamma_{{\bf p}}-A_{2})-\lambda_{2}
(2\chi^{z}_{2}\gamma_{{\bf p }}'-\chi_{2})$, $\lambda_{1}=
2ZJ_{eff}$, $\lambda_{2}=4Z\phi_{2} t'$,  $A_{1}= \epsilon
\chi^{z}_{1}+\chi_{1}/2$, $A_{2} =\chi^{z}_{1}+\epsilon
\chi_{1}/2$, $\epsilon=1+2t\phi_{1} /J_{{\rm eff}}$, the dressed
holon's particle-hole parameters $\phi_{1}=\langle
h^{\dagger}_{i\sigma}h_{i+\hat{\eta}\sigma}\rangle$ and $\phi_{2}=
\langle h^{\dagger}_{i\sigma} h_{i+\hat{\tau}\sigma}\rangle$, the
spin correlation functions $\chi^{z}_{1}=\langle S_{i}^{z}
S_{i+\hat{\eta}}^{z}\rangle$ and $\chi^{z}_{2}=\langle S_{i}^{z}
S_{i+\hat{\tau}}^{z}\rangle$, and the MF dressed spin excitation
spectrum,
\begin{eqnarray}
\omega^{2}_{{\bf p}}&=& \lambda_{1}^{2}[(A_{4}-\alpha\epsilon
\chi^{z}_{1}\gamma_{{\bf p}}-{1\over 2Z}\alpha\epsilon\chi_{1})
(1-\epsilon\gamma_{{\bf p}})\nonumber \\
&+&{1\over 2}\epsilon(A_{3}-{1\over 2} \alpha\chi^{z}_{1}-\alpha
\chi_{1}\gamma_{{\bf p}})(\epsilon-\gamma_{{\bf p}})] \nonumber \\
&+&\lambda_{2}^{2}[\alpha(\chi^{z}_{2}\gamma_{{\bf p}}'-{3\over
2Z}\chi_{2})\gamma_{{\bf p}}'+{1\over 2}(A_{5}-{1\over 2}\alpha
\chi^{z}_{2})]\nonumber \\
&+&\lambda_{1}\lambda_{2}[\alpha\chi^{z}_{1}(1-\epsilon
\gamma_{{\bf p}})\gamma_{{\bf p}}'+{1\over 2}
\alpha(\chi_{1}\gamma_{{\bf p}}'-C_{3})(\epsilon-
\gamma_{{\bf p}})\nonumber \\
&+&\alpha \gamma_{{\bf p}}'(C^{z}_{3}-\epsilon \chi^{z}_{2}
\gamma_{{\bf p}})-{1\over 2}\alpha\epsilon(C_{3}- \chi_{2}
\gamma_{{\bf p}})],
\end{eqnarray}
with $A_{3}=\alpha C_{1}+(1-\alpha)/(2Z)$, $A_{4}=\alpha C^{z}_{1}
+(1-\alpha)/(4Z)$, $A_{5}=\alpha C_{2}+(1-\alpha)/(2Z)$, and the
spin correlation functions
$C_{1}=(1/Z^{2})\sum_{\hat{\eta},\hat{\eta'}}\langle
S_{i+\hat{\eta}}^{+}S_{i+\hat{\eta'}}^{-}\rangle$,
$C^{z}_{1}=(1/Z^{2})\sum_{\hat{\eta},\hat{\eta'}}\langle
S_{i+\hat{\eta}}^{z}S_{i+\hat{\eta'}}^{z}\rangle$,
$C_{2}=(1/Z^{2})\sum_{\hat{\tau},\hat{\tau'}}\langle
S_{i+\hat{\tau}}^{+}S_{i+\hat{\tau'}}^{-}\rangle$,
$C_{3}=(1/Z)\sum_{\hat{\tau}}\langle S_{i+\hat{\eta}}^{+}
S_{i+\hat{\tau}}^{-}\rangle$, and $C^{z}_{3}=(1/Z)
\sum_{\hat{\tau}}\langle S_{i+\hat{\eta}}^{z}
S_{i+\hat{\tau}}^{z}\rangle$. In order to satisfy the sum rule of
the correlation function $\langle S^{+}_{i}S^{-}_{i}\rangle=1/2$
in the case without AFLRO, the important decoupling parameter
$\alpha$ has been introduced in the MF calculation
\cite{feng2,feng4}, which can be regarded as the vertex
correction.

The self-energy function $\Sigma^{(h)}_{2}(k)$ describes the
effective dressed holon gap function, since both doping and
temperature dependence of the pairing force and dressed holon gap
function have been incorporated into $\Sigma^{(h)}_{2}(k)$, while
the self-energy function $\Sigma^{(h)}_{1}(k)$ renormalizes the MF
dressed holon spectrum, and therefore it describes the
quasiparticle coherence. Moreover, $\Sigma^{(h)}_{2}(k)$ is an
even function of $i\omega_{n}$, while $\Sigma^{(h)}_{1}(k)$ is
not. For the convenience, $\Sigma^{(h)}_{1}(k)$ can be broken up
into its symmetric and antisymmetric parts as, $\Sigma^{(h)}_{1}
(k)=\Sigma^{(h)}_{1e} (k)+i\omega_{n} \Sigma^{(h)}_{1o}(k)$, then
both $\Sigma^{(h)}_{1e} (k)$ and $\Sigma^{(h)}_{1o}(k)$ are even
functions of $i\omega_{n}$. In this case, the quasiparticle
coherent weight can be defined as $Z^{-1}_{F}(k) =1-
\Sigma^{(h)}_{1o}(k)$. As in the conventional superconductor
\cite{eliashberg}, the retarded function ${\rm Re}
\Sigma^{(h)}_{1e}(k)$ is a constant, independent of (${\bf k},
\omega$), and it just renormalizes the chemical potential,
therefore it can be dropped. Furthermore, we only study the static
limit of the effective dressed holon gap function and
quasiparticle coherent weight, i.e., $\Sigma^{(h)}_{2}(k)=
\bar{\Delta}_{h}({\bf k})$, and $Z^{-1}_{F}({\bf k})=1-
\Sigma^{(h)}_{1o}({\bf k})$. Although $Z_{F}({\bf k})$ still is a
function of ${\bf k}$, the wave vector dependence is unimportant,
since everything happens at the electron Fermi surface. As in the
previous discussions within the $t$-$J$ model \cite{feng3}, the
special wave vector can be estimated qualitatively from the
electron momentum distribution as ${\bf k}_{0}={\bf k_{A}}-{\bf
k_{F}}$ with ${\bf k_{A}}= [\pi, \pi]$ and ${\bf k_{F}}\approx
[(1-x)\pi/2, (1-x)\pi/2]$, which guarantees $Z_{F}=Z_{F} ({\bf
k}_{0})$ near the electron Fermi surface. In this case, the
dressed holon diagonal and off-diagonal Green's functions in Eqs.
(4a) and (4b) can be expressed explicitly as,
\begin{mathletters}
\begin{eqnarray}
g(k)&=&Z_{F}{U^{2}_{h{\bf k}}\over i\omega_{n}-E_{h{\bf k}}}
+Z_{F}{V^{2}_{h{\bf k}}\over i\omega_{n}+E_{h{\bf k}}}, \\
\Im^{\dagger}(k)&=&-Z_{F}{\bar{\Delta}_{hZ}({\bf k})\over
2E_{h{\bf k}}}\left ( {1\over i\omega_{n}-E_{h{\bf k}}}- {1\over
i\omega_{n}+ E_{h{\bf k}}}\right ),
\end{eqnarray}
\end{mathletters}
with the dressed holon quasiparticle coherence factors
$U^{2}_{h{\bf k}}=(1+\bar{\xi_{{\bf k}}}/E_{h{\bf k}})/2$ and
$V^{2}_{h{\bf k}}=(1-\bar{\xi_{{\bf k}}}/E_{h{\bf k}})/2$,
$\bar{\xi_{{\bf k}}}=Z_{F}\xi_{{\bf k}}$, $\bar{\Delta}_{hZ}({\bf
k})=Z_{F}\bar{\Delta}_{h}({\bf k})$, and the dressed holon
quasiparticle spectrum $E_{h{\bf k}}= \sqrt {\bar{\xi^{2}_{{\bf
k}}}+\mid\bar{\Delta}_{hZ}({\bf k}) \mid^{2}}$.

Experimentally, some results seem consistent with an s-wave
pairing \cite{chaudhari}, while other measurements gave the
evidence in favor of the d-wave pairing \cite{martindale,tsuei}.
These experiments reflect a fact that the d-wave gap function
$\propto {\rm k}^{2}_{x}-{\rm k}^{2}_{y}$ belongs to the same
representation $\Gamma_{1}$ of the orthorhombic crystal group as
does s-wave gap function $\propto {\rm k}^{2}_{x}+{\rm
k}^{2}_{y}$. Within the $t$-$J$ model, we \cite{feng3} have shown
that the electron Cooper pairs have a dominated d-wave symmetry
over a wide range of the doping concentration, around the optimal
doping. To make the discussion simpler, we only consider the
d-wave case, i.e., $\bar{\Delta}_{hZ} ({\bf k})=\bar{\Delta}_{hZ}
\gamma^{(d)}_{{\bf k}}$, with $\gamma^{(d)}_{{\bf k}}=({\rm cos}
k_{x}-{\rm cos}k_{y})/2$. In this case, the dressed holon
effective gap parameter and quasiparticle coherent weight in Eqs.
(5a) and (5b) satisfy following two equations,
\begin{mathletters}
\begin{eqnarray}
&1&={1\over N^{3}}\sum_{{\bf k,q,p}}(Zt\gamma_{{\bf k+q}}-Zt'
\gamma_{{\bf k+q}}')^{2}\gamma^{(d)}_{{\bf k-p+q}}
\gamma^{(d)}_{{\bf k}}{Z^{2}_{F}\over E_{h{\bf k}}}{B_{{\bf q}}
B_{{\bf p}}\over \omega_{{\bf q}}\omega_{{\bf p}}}\nonumber \\
&\times&\left({F^{(1)}_{1}({\bf k,q,p}) \over (\omega_{{\bf p}}
-\omega_{{\bf q}})^{2}-E^{2}_{h{\bf k}}}-{F^{(2)}_{1} ({\bf k,
q,p})\over (\omega_{{\bf p}}+\omega_{{\bf q}})^{2}-
E^{2}_{h{\bf k}}}\right ) ,\\
&Z^{-1}_{F}&=1+{1\over N^{2}}\sum_{{\bf q,p}}(Zt\gamma_{{\bf
p+k_{0}}}-Zt'\gamma_{{\bf p+k_{0}}}')^{2}Z_{F}{B_{{\bf q}} B_{{\bf
p}}\over 4\omega_{{\bf q}}\omega_{{\bf p}}}\nonumber \\
&\times&\left({F^{(1)}_{2} ({\bf q,p})\over (\omega_{{\bf p}}-
\omega_{{\bf q}}-E_{h{\bf p-q+k_{0}}})^{2}}+{F^{(2)}_{2}({\bf
q,p})\over (\omega_{{\bf p}}- \omega_{{\bf q}}+E_{h{\bf
p-q+k_{0}}})^{2}}\right .\nonumber \\
&+&\left .{F^{(3)}_{2}({\bf q,p })\over (\omega_{{\bf p}}+
\omega_{{\bf q} }-E_{h{\bf p-q+k_{0}}})^{2}}+{F^{(4)}_{2}({\bf
q,p})\over (\omega_{{\bf p}}+ \omega_{{\bf q}}+E_{h{\bf p-
q+k_{0}}})^{2}} \right ) ,
\end{eqnarray}
\end{mathletters}
respectively, where $F^{(1)}_{1}({\bf k,q,p})= (\omega_{{\bf p}}-
\omega_{{\bf q}}) [n_{B}(\omega_{{\bf q}})- n_{B}(\omega_{{\bf p}
})][1-2 n_{F}(E_{h{\bf k}})]+E_{h{\bf k}}[n_{B}(\omega_{{\bf p}})
n_{B}(-\omega_{{\bf q}})+n_{B}(\omega_{{\bf q}})n_{B}
(-\omega_{{\bf p}})]$, $F^{(2)}_{1}({\bf k,q,p})=(\omega_{{\bf p}
}+\omega_{{\bf q}})[n_{B}(-\omega_{{\bf p}})-n_{B}(\omega_{{\bf q
}})][1-2 n_{F}(E_{h{\bf k}})]+E_{h{\bf k}}[n_{B}(\omega_{{\bf p}})
n_{B}(\omega_{{\bf q}})+n_{B}(-\omega_{{\bf p}})n_{B}(-
\omega_{{\bf q}})]$, $F^{(1)}_{2}({\bf q,p})=n_{F}(E_{h{\bf
p-q+k_{0}}})[n_{B}(\omega_{{\bf q}})-n_{B}(\omega_{{\bf p}})]
-n_{B}(\omega_{{\bf p}})n_{B}(-\omega_{{\bf q}})$, $F^{(2)}_{2}
({\bf q,p})=n_{F}(E_{h{\bf p-q+k_{0}}})[n_{B}(\omega_{{\bf p}})
-n_{B}(\omega_{{\bf q}})]-n_{B}(\omega_{{\bf q}})n_{B}
(-\omega_{{\bf p}})$, $F^{(3)}_{2}({\bf q,p})= n_{F}(E_{h{\bf
p-q+k_{0}}})[n_{B}(\omega_{{\bf q}})-n_{B}(-\omega_{{\bf p}})]
+n_{B}(\omega_{{\bf p}})n_{B} (\omega_{{\bf q}})$, and
$F^{(4)}_{2}({\bf q,p})=n_{F}(E_{h{\bf p-q+k_{0}}})
[n_{B}(-\omega_{{\bf q}})-n_{B} (\omega_{{\bf p}})]+ n_{B}
(-\omega_{{\bf p}})n_{B}(-\omega_{{\bf q}})$. These two equations
must be solved simultaneously with other self-consistent equations
\cite{feng2}, then all order parameters, decoupling parameter
$\alpha$, and chemical potential $\mu$ are determined by the
self-consistent calculation. With the above discussions, we now
can obtain the dressed holon pair gap function in terms of the
off-diagonal Green's function (7b) as $\Delta_{h}({\bf k})=-
(1/\beta) \sum_{i\omega_{n}}\Im^{\dagger}({\bf k},i\omega_{n})$,
then the dressed holon pair order parameter can be evaluated as,
\begin{eqnarray}
\Delta_{h}={2\over N}\sum_{{\bf k}} [\gamma^{(d)}_{{\bf k}} ]^{2}
{Z_{F}\bar{\Delta}_{hZ}\over E_{h{\bf k}}}{\rm tanh} [{1\over 2}
\beta E_{h{\bf k}}].
\end{eqnarray}
This dressed holon pairing state originating from the kinetic
energy terms by exchanging dressed spin excitations also leads to
form the electron Cooper pairing state \cite{feng2}, and the SC
gap function is obtained from the electron off-diagonal Green's
function $\Gamma^{\dagger}(i-j,t-t')=\langle \langle
C^{\dagger}_{i\uparrow}(t);C^{\dagger}_{j\downarrow}(t')
\rangle\rangle$, which is a convolution of the dressed spin
Green's function and dressed holon off-diagonal Green's function
and reflects the charge-spin recombination \cite{anderson2}. In
the present case, this electron off-diagonal Green's function can
be evaluated in terms of the MF dressed spin Green's function and
dressed holon off-diagonal Green's function (7b) as,
\begin{eqnarray}
\Gamma^{\dagger}(k)&=&{1\over N}\sum_{{\bf p}}
Z_{F}{\bar{\Delta}^{(a)}_{hZ}({\bf p+k})\over 2E_{h{\bf p+k}}}
{B_{{\bf p }}\over 2\omega_{{\bf p}}}\left \{F^{(1)}_{3}({\bf k,p}
) \right .\nonumber \\
&\times&\left ({1\over i\omega_{n}-E_{h{\bf p+k}}-\omega_{{\bf p}
}}-{1\over i\omega_{n}+E_{h{\bf p+k}}+\omega_{{\bf p}}}\right )
\nonumber \\
&-& F^{(2)}_{3}({\bf k,p})\left ({1\over i\omega_{n}+
E_{h{\bf p+k}}-\omega_{{\bf p}}}\right .\nonumber \\
&-&\left . \left . {1\over i\omega_{n}-E_{h{\bf p+k} }+
\omega_{{\bf p}}} \right ) \right \},
\end{eqnarray}
with $F^{(1)}_{3}({\bf k,p})=1-n_{F}(E_{h{\bf k+p}})+n_{B}
(\omega_{{\bf p}})$ and $F^{(2)}_{3}({\bf k,p})=n_{F}(E_{h{\bf
k+p}})+ n_{B}(\omega_{{\bf p}})$, then the SC gap function is
obtained from the above electron off-diagonal Green's function as,
\begin{eqnarray}
\Delta({\bf k})&=&-{1\over N}\sum_{{\bf p}}{Z_{F}
\bar{\Delta}_{Zh} ({\bf p-k})\over 2E_{h{\bf p-k}}}{\rm tanh}
[{1\over 2}\beta E_{h{\bf p-k}}]\nonumber \\
&\times&{B_{{\bf p}}\over 2\omega_{{\bf p}}}{\rm coth}[{1\over 2}
\beta\omega_{{\bf p}}].
\end{eqnarray}
From this SC gap function, the SC gap parameter in Eq. (3) is
obtained as $\Delta=-\chi_{1} \Delta_{h}$. Since both dressed
holon (then electron) pairing gap parameter and pairing
interaction in cuprate superconductors are doping dependent,
therefore the experimental observed SC gap parameter should be an
effective SC gap parameter $\bar{\Delta}\sim -\chi_{1}\bar
{\Delta}_{h}$, which measures the strength of the binding of
electrons into electron Cooper pairs. In Fig. 1, we plot the
effective dressed holon pairing (a) and effective SC (b) gap
parameters in the d-wave symmetry as a function of the hole doping
concentration at $T=0.002J$ for $t/J=2.5$ and $t'/J=0.3$ (solid
line) and $t/J=2.5$ and $t'=0$ (dashed line). For comparison, the
experimental result \cite{wen} of the upper critical field as a
function of the hole doping concentration is also shown in Fig.
1(b). In a given doping concentration, the upper critical field is
defined as the critical field that destroys the SC-state at the
zero temperature, therefore the upper critical field also measures
the strength of the binding of electrons into Cooper pairs like
the effective SC gap parameter \cite{wen}. In other words, both
effective SC gap parameter and upper critical field have a similar
doping dependence \cite{wen}. In this sense, our result is in good
agreement with the experimental data \cite{wen}. Our result also
shows that the effect of $t'$ on the SC-state of the $t$-$J$ model
is to enhance the amplitude of the effective dressed holon (then
electron) pairing gap parameter, and shift the maximal value of
$\bar{\Delta}_{h}$ (then $\bar{\Delta}$) towards to the low doping
regime. In particular, the value of $\bar{\Delta}$ in the
$t$-$t'$-$J$ model increases with increasing doping in the
underdoped regime, and reaches a maximum in the optimal doping
$x_{{\rm opt}}\approx 0.15$, then decreases in the overdoped
regime. Since the effective dressed holon pairing gap parameter
measures the strength of the binding of dressed holons into
dressed holon pairs, then our results also show that although the
superconductivity is driven by the kinetic energy by exchanging
dressed spin excitations, the strength of the binding of electrons
into electron Cooper pairs is still suppressed by AFSRC. Based on
the numerical simulations, it has been shown \cite{shih} that the
SC correlation of the $t$-$J$ model is enhanced by introducing
$t'$, where the particular correlation between the SC gap and
electron occupation at $[\pi,0]$ point is the main reason for
enhancement of pairs, which is consistent with our present result.
However, their result also shows \cite{shih} that the SC
correlation becomes strongest shifts to the overdoped regime by
introducing $t'$, and therefore the SC correlation is greatly
enhanced in the overdoped regime, which is inconsistent with our
present result. The reason for this inconsistency is not clear,
and the related issue is under investigation now.

\begin{figure}[prb]
\epsfxsize=3.5in\centerline{\epsffile{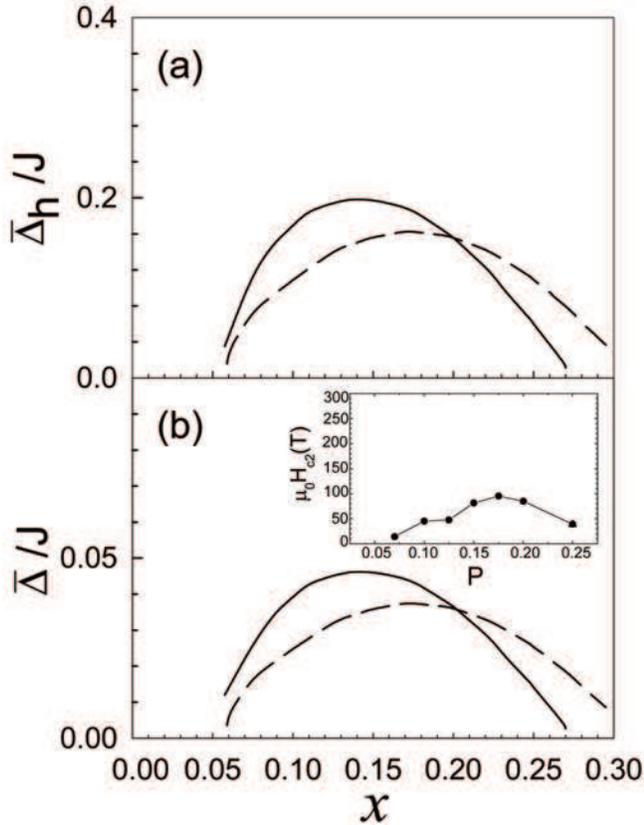}}\caption{ The
effective dressed holon pairing (a) and effective superconducting
(b) gap parameters in the d-wave symmetry as a function of the
hole doping concentration in $T=0.002J$ for $t/J=2.5$ and
$t'/t=0.3$ (solid line) and $t/J=2.5$ and $t'=0$ (dashed line).
Inset: the experimental result of the upper critical field as a
function of the hole doping concentration taken from Ref. [24].}
\end{figure}

\begin{figure}[prb]
\epsfxsize=3.5in\centerline{\epsffile{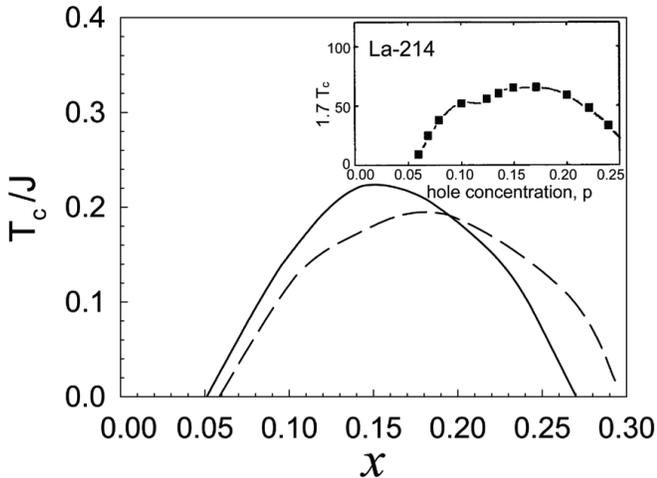}} \caption{ The
superconducting transition temperature as a function of the hole
doping concentration in the d-wave symmetry for $t/J=2.5$ and
$t'/t=0.3$ (solid line) and $t/J=2.5$ and $t'=0$ (dashed line).
Inset: the experimental result taken from Ref. [11].}
\end{figure}

Now we turn to discuss the effect of $t'$ on the SC transition
temperature. As in the case of the $t$-$J$ model \cite{feng2}, the
SC transition temperature $T_{c}$ occurring in the case of the SC
gap parameter $\Delta=0$ in Eq. (11) is identical to the dressed
holon pair transition temperature occurring in the case of the
effective holon pairing gap parameter $\bar{\Delta}_{hZ} =0$. In
this case, we have performed a calculation for the doping
dependence of the SC transition temperature, and the result of
$T_{c}$ as a function of the hole doping concentration in the
d-wave symmetry for $t/J=2.5$ and $t'/J=0.3$ (solid line) and
$t/J=2.5$ and $t'=0$ (dashed line) is plotted in Fig. 2 in
comparison with the experimental result \cite{tallon} (inset). Our
result shows that the maximal SC transition temperature T$_{c}$ of
the $t$-$t'$-$J$ model occurs around the optimal doping $x_{{\rm
opt}}\approx 0.15$, and then decreases in both underdoped and
overdoped regimes. Furthermore, T$_{c}$ in the underdoped regime
is proportional to the hole doping concentration $x$, and
therefore T$_{c}$ in the underdoped regime is set by the hole
doping concentration \cite{uemura}. This reflects that the density
of the dressed holons directly determines the superfluid density
in the underdoped regime. Using an reasonably estimative value of
$J\sim 800$K to 1200K in doped cuprates, the SC transition
temperature in the optimal doping is T$_{c}\approx 0.22J \approx
176{\rm K}\sim 264{\rm K}$, in qualitative agreement with the
experimental data \cite{tallon}. In comparison with the result of
the $t$-$J$ model \cite{feng3}, our present result also shows that
$t'$ plays an important role in enhancing the SC transition
temperature of the $t$-$J$ model and in shifting the maximal value
of $T_{c}$ towards to the low doping regime.

\begin{figure}[prb]
\epsfxsize=3.5in\centerline{\epsffile{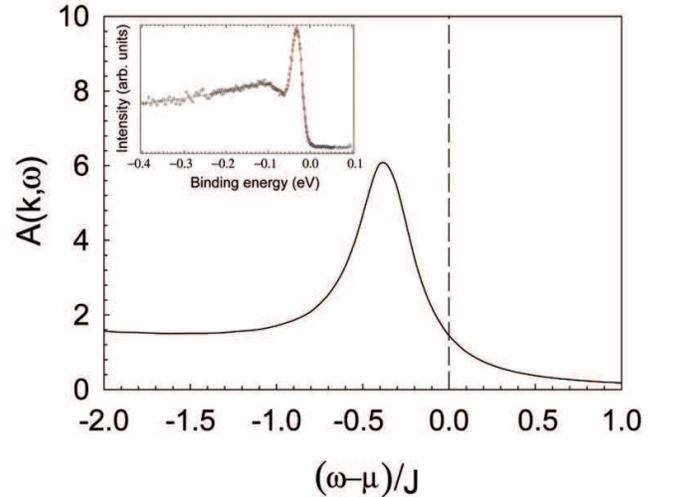}}\caption{ The
electron spectral function with the d-wave symmetry at $[\pi,0]$
point in $x_{{\rm opt}}=0.15$ and $T=0.002J$ for $t/J=2.5$ and
$t'/J=0.3$.  Inset: the experimental result taken from Ref. [29].
}
\end{figure}

For cuprate superconductors, ARPES experiments have produced some
interesting data that introduce important constraints on the SC
theory \cite{shen2}. Since cuprates superconductors are highly
anisotropic materials, therefore the electron spectral function
$A({\bf k}, \omega)$ is dependent on the in-plane momentum
\cite{shen2}. Although the electron spectral function in doped
cuprates obtained from ARPES is very broad in the normal-state,
indicating that there are no quasiparticles \cite{shen2}. However,
in the SC-state, the full energy dispersion of quasiparticles has
been observed \cite{ding2}. According to a comparison of the
density of states as measured by scanning tunnelling microscopy
\cite{dewilde} and ARPES spectral function \cite{ding,shen2} at
$[\pi,0]$ point on identical samples, it has been shown that the
most contributions of the electron spectral function come from
$[\pi,0]$ point. In addition, the d-wave gap, and therefore the
electron pairing energy scale, is maximized at $[\pi,0]$ point.
Although the sharp SC quasiparticle peak at $[\pi,0]$ point in
cuprate superconductors has been widely studied, the orgin and its
implications are still under debate \cite{ding2}. As a test of the
kinetic energy driven superconductivity in doped cuprates
\cite{feng2}, we now study this issue. For discussions of the
electron spectral function, we need to calculate the electron
diagonal Green's function $G(i-j,t-t')= \langle\langle C_{i\sigma}
(t);C^{\dagger}_{j\sigma}(t') \rangle \rangle$, which is a
convolution of the dressed spin Green's function and dressed holon
diagonal Green's function, and can be evaluated in terms of the MF
dressed spin Green's function $D^{(0)}(p)$ and dressed holon
diagonal Green's function $g(k)$ in Eq. (7a) as,
\begin{eqnarray}
G({\bf k},\omega)&=&{1\over N}\sum_{{\bf p}}Z_{F}{B_{{\bf p}}\over
2\omega_{{\bf p}}}\left \{\left ({U^{2}_{h{\bf p+k}}\over\omega +
E_{h{\bf p+k}}-\omega_{{\bf p}}}\right .\right . \nonumber \\
&+&\left .{V^{2}_{h{\bf p+k}}\over\omega-E_{h{\bf p+k}}+
\omega_{{\bf p}}}\right )[n_{F}(E_{h{\bf p+k}})+n_{B}
(\omega_{{\bf p}})]\nonumber \\
&+&[1-n_{F}(E_{h{\bf p+k}})+n_{B}(\omega_{{\bf p}})]\nonumber\\
&\times& \left .\left ({U^{2}_{h{\bf p+k}}\over\omega+E_{h{\bf
p+k}}+\omega_{{\bf p}}} +{V^{2}_{h{\bf p+k}}\over\omega-E_{h{\bf
p+k}}-\omega_{{\bf p}}}\right ) \right \} ,
\end{eqnarray}
then from this electron diagonal Green's function, the electron
spectral function $A({\bf k},\omega)=-2{\rm Im} G({\bf k},\omega)$
is obtained as,
\begin{eqnarray}
A({\bf k},\omega)&=&{1\over N}\sum_{{\bf p}}Z_{F}{B_{{\bf p}}\over
2\omega_{{\bf p}}}\left \{[n_{F}(E_{h{\bf p+k}})+n_{B}
(\omega_{{\bf p}})]\right .\nonumber \\
&\times&[U^{2}_{h{\bf p+k}}\delta(\omega+E_{h{\bf p+k}}
-\omega_{{\bf p}})\nonumber \\
&+& V^{2}_{h{\bf p+k}}\delta(\omega-E_{h{\bf p+k}}+ \omega_{{\bf
p}})] \nonumber \\
&+&[1-n_{F}(E_{h{\bf p+k}})+n_{B}(\omega_{{\bf p}})] \nonumber
\\
&\times&[U^{2}_{h{\bf p+k}}\delta(\omega+E_{h{\bf p+k}}
+\omega_{{\bf p}})\nonumber \\
&+& \left . V^{2}_{h{\bf p+k}}\delta(\omega-E_{h{\bf p+k}}
-\omega_{{\bf p}}) ]\right \} .
\end{eqnarray}
We have performed the calculation for this electron spectral
function, and the result of $A({\bf k},\omega)$ at $[\pi,0]$ point
in the optimal doping $x_{{\rm opt}}=0.15$ with $T=0.002J$ for
$t/J=2.5$ and $t'/J=0.3$ is plotted in Fig. 3 in comparison with
the experimental result \cite{ding} (inset). Our result shows that
there is a sharp SC quasiparticle peak near the electron Fermi
surface at $[\pi,0]$ point, and the position of this SC
quasiparticle peak is located at $\omega_{{\rm peak}}\approx 0.4J
\approx 0.028$eV$\sim 0.04$eV, which is quantitatively consistent
with the $\omega_{{\rm peak}} \approx 0.03$eV observed \cite{ding}
in the cuprate superconductor Bi$_{2}$Sr$_{2}$CaCu$_{2}$O$_{8+x}$.
Our result also shows that the dressed holon pairs condense with
the d-wave symmetry in a wide range of the doping concentration,
then the electron Cooper pairs originating from the dressed holon
pairing state are due to the charge-spin recombination, and their
condensation automatically gives the electron quasiparticle
character. Furthermore, we have discussed the temperature
dependence of the electron spectral function and overall
quasiparticle dispersion, and these and related theoretical
results will be presented elsewhere.

Our present result also indicates that the SC-state of cuprate
superconductors is the conventional BCS like \cite{bcs}, this can
be understood from the electron diagonal and off-diagonal Green's
functions in Eqs. (12) and (10), which can be rewritten as,
\begin{mathletters}
\begin{eqnarray}
G({\bf k},\omega)&=&{1\over N}\sum_{{\bf p}}Z_{F}{B_{{\bf p}}\over
4\omega_{{\bf p}}}\left \{\left ( {U^{2}_{h{\bf p+k}}\over\omega +
E_{h{\bf p+k}}-\omega_{{\bf p}}}\right .\right .\nonumber \\
&+&{U^{2}_{h{\bf p+k}}\over\omega+E_{h{\bf p+k}}+\omega_{{\bf p}}}
+{V^{2}_{h{\bf p+k}}\over\omega-E_{h{\bf p+k}}+\omega_{{\bf p}}}
\nonumber \\
&+&\left . {V^{2}_{h{\bf p+k}}\over\omega-E_{h{\bf p+k}} -
\omega_{{\bf p}}}\right){\rm coth}[{1\over 2}\beta\omega_{{\bf p}
}]\nonumber \\
&+&{\rm tanh}[{1\over 2}\beta E_{h{\bf p+k}}]\left ({U^{2}_{h{\bf
p+k}}\over\omega+E_{h{\bf p+k}}+\omega_{{\bf p}}}\right .
\nonumber \\
&-&{U^{2}_{h{\bf p+k}}\over\omega+E_{h{\bf p+k}}-\omega_{{\bf p}}}
+{V^{2}_{h{\bf p+k}}\over\omega-E_{h{\bf p+k}}- \omega_{{\bf p}}}
\nonumber \\
&-&\left . \left . {V^{2}_{h{\bf p+k}}\over\omega-E_{h{\bf p+k}}
+\omega_{{\bf p}}} \right ) \right \} ,\\
\Gamma^{\dagger}({\bf k},\omega)&=&{1\over N}\sum_{{\bf p}}
Z_{F}{\bar{\Delta}_{hZ}({\bf p+k})\over 2E_{h{\bf p+k}}}{B_{{\bf p
}}\over 4\omega_{{\bf p}}}\left \{{\rm coth}[{1\over 2}\beta
\omega_{{\bf p}}]\right . \nonumber \\
&\times& \left ({1\over\omega-E_{h{\bf p+k}}-\omega_{{\bf p}}}+
{1\over \omega-E_{h{\bf p+k}}+\omega_{{\bf p}}}\right . \nonumber
\\
&-& \left . {1\over\omega+E_{h{\bf p+k}}+\omega_{{\bf p}}}
-{1\over\omega+E_{h{\bf p+k}}-\omega_{{\bf p}}}\right )\nonumber
\\
&+&{\rm tanh}[{1\over 2}\beta E_{h{\bf p+k}}]\left ({1\over\omega-
E_{h{\bf p+k}}-\omega_{{\bf p}}}\right .\nonumber \\
&-&{1\over\omega-E_{h{\bf p+k}}+\omega_{{\bf p}}}-{1\over \omega
+E_{h{\bf p+k}}+\omega_{{\bf p}}}\nonumber \\
&+&\left .\left . {1\over \omega +E_{h{\bf p+k}}-\omega_{{\bf p}}}
\right )\right \},
\end{eqnarray}
\end{mathletters}
respectively. Since the dressed spins center around
$[\pm\pi,\pm\pi]$ in the Brillouin zone in the mean-field level
\cite{feng1,feng2,feng3}, therefore the above electron diagonal
and off-diagonal Green's functions can be approximately reduced in
terms of $\omega_{{\bf p}=\pm\pi,\pm\pi}\sim 0$ and the equation
\cite{feng1,feng2} $1/2=\langle S_{i}^{z}S_{i}^{z}\rangle=1/N
\sum_{{\bf p}}B_{{\bf p} }{\rm coth}(\beta\omega_{{\bf p}}/2)
/(2\omega_{{\bf p}})$ as,
\begin{mathletters}
\begin{eqnarray}
g({\bf k},\omega)&\propto&Z_{F}{V^{2}_{{\bf k}}\over
\omega-E_{{\bf k}}}
+Z_{F}{U^{2}_{{\bf k}}\over \omega+E_{{\bf k}}},\\
\Gamma^{\dagger}({\bf k},\omega)&\propto& Z_{F}{\bar{\Delta}_{hZ}
({\bf k })\over 2E_{{\bf k}}}\left ({1\over \omega-E_{{\bf k}}}+
{1\over \omega+E_{{\bf k}}}\right ),
\end{eqnarray}
\end{mathletters}
with the electron quasiparticle coherence factors $U^{2}_{{\bf k}}
\propto V^{2}_{{\bf k+k_{A}}}$ and $V^{2}_{{\bf k}}\propto
U^{2}_{{\bf k+k_{A}} }$, and electron quasiparticle spectrum
$E_{{\bf k}} \propto E_{h{\bf k+k_{A}}}$, with ${\bf k_{A}}=
[\pi,\pi]$, {i.e.}, the hole-like dressed holon quasiparticle
coherence factors $U_{h{\bf k}}$ and $V_{h{\bf k}}$ have been
transferred into the electron quasiparticle coherence factors
$U_{{\bf k}}$ and $V_{{\bf k}}$ by the convolution of the dressed
spin Green's function and dressed holon diagonal Green's function
due to the charge-spin recombination, this is why the basic BCS
formalism \cite{ding2} is still valid in discussions of the doping
dependence of the effective SC gap parameter and SC transition
temperature, and electron spectral function \cite{ding2}, although
the pairing mechanism is driven by the kinetic energy by
exchanging dressed spin excitations, and other exotic properties
are beyond the BCS theory.

\begin{figure}[prb]
\epsfxsize=3.5in\centerline{\epsffile{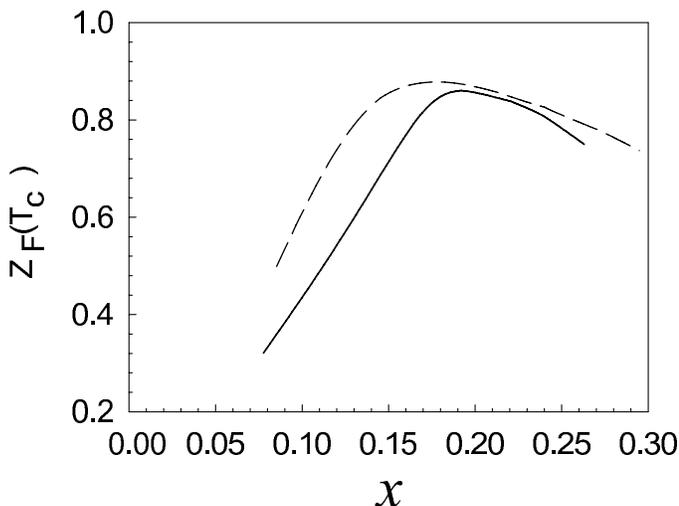}}\caption{ The
quasiparticle coherent weight $Z_{F}(T_{c})$ as a function of the
hole doping concentration for $t/J=2.5$ and $t'/t=0.3$ (solid
line) and $t/J=2.5$ and $t'=0$ (dashed line). }
\end{figure}

The essential physics of superconductivity in the present
$t$-$t'$-$J$ model is the same as that in the $t$-$J$ model
\cite{feng2,feng3}. The antisymmetric part of the self-energy
function $\Sigma^{(h)}_{1o}({\bf k})$ (then $Z_{F}$) describes the
dressed holon (then electron) quasiparticle coherence, and
therefore $Z_{F}$ is closely related to the SC quasiparticle
density, while the self-energy function $\Sigma^{(h)}_{2}({\bf k})
$ describes the effective dressed holon (then electron) pairing
gap function. In particular, both $Z_{F}$ and $\Sigma^{(h)}_{2}
({\bf k})$ are doping and temperature dependent. Since the
SC-order is established through an emerging quasiparticle
\cite{ding,he}, therefore the SC-order is controlled by both gap
function and quasiparticle coherence, and is reflected explicitly
in the self-consistent equations (8a) and (8b). The dressed holons
(then electrons) interact by exchanging the dressed spins and that
this interaction is attractive. This attractive interaction leads
to form the dressed holon pairs (then electron Cooper pairs). The
perovskite parent compound of doped cuprate superconductors is a
Mott insulator, when holes are doped into this insulator, there is
a gain in the kinetic energy per hole proportional to $t$ due to
hopping, but at the same time, the spin correlation is destroyed,
costing an energy of approximately $J$ per site, therefore the
doped holes into the Mott insulator can be considered as a
competition between the kinetic energy ($xt$) and magnetic energy
($J$), and the magnetic energy decreases with increasing doping.
In the underdoped and optimally doped regimes, the magnetic energy
is rather too large, and the dressed holon (then electron)
attractive interaction by exchanging the dressed spin is also
rather strong to form the dressed holon pairs (then electron
Cooper pairs) for the most dressed holons (then electrons),
therefore the number of the dressed holon pairs (then electron
Cooper pairs), SC transition temperature \cite{uemura}, and
quasiparticle coherent weight \cite{ding,he} are proportional to
the hole doping concentration. However, in the overdoped regime,
the magnetic energy is relatively small, and the dressed holon
(then electron) attractive interaction by exchanging the dressed
spin is also relatively weak, in this case, not all dressed holons
(then electrons) can be bounden as dressed holon pairs (then
electron Cooper pairs) by the weak attractive interaction, and
therefore the number of the dressed holon pairs (then electron
Cooper pairs), SC transition temperature \cite{tallon}, and
quasiparticle coherent weight \cite{he} decrease with increasing
doping. To show this point clearly, we plot the quasiparticle
coherent weight $Z_{F}(T_{c})$ as a function of the hole doping
concentration for $t/J=2.5$ and $t'/t=0.3$ (solid line) and
$t/J=2.5$ and $t'=0$ (dashed line) in Fig. 4. As seen from Fig. 4,
the doping dependent behavior of the quasiparticle coherent weight
resembles that of the superfluid density in cuprate
superconductors, i.e., $Z_{F}(T_{c})$ grows linearly with the hole
doping concentration in the underdoped and optimally doped
regimes, and then decreases with increasing doping in the
overdoped regime, which leads to that the SC transition
temperature reaches a maximum in the optimal doping, and then
decreases in both underdoped and overdoped regimes. The behavior
of the doping dependence of $Z_{F}$ in Fig. 4 is consistent with
the experimental result \cite{ding,he}, where the quasiparticle
coherent weight increases monotonically with increasing doping in
the underdoped and optimally doped regimes \cite{ding}, and then
decreases with increasing doping in the overdoped regime
\cite{he}. On the other hand, the electronic structure becomes
asymmetric and hole doping shifts the Fermi surface to the van
Hove singularity when the additional second neighbor hopping $t'$
is introduced in the $t$-$J$ model \cite{newns}, which leads to
increase the density of states at the Fermi energy, then the SC
correlation is enhanced. Furthermore, the additional second
neighbor hopping $t'$ in the $t$-$J$ model is equivalent to
increase the kinetic energy. These are also why $t'$ plays an
important role in enhancing the SC transition temperature of the
$t$-$J$ model under the kinetic energy driven SC mechanism.

In summary, we have discussed the effect of the additional second
neighbor hopping $t'$ on the SC-state of the $t$-$J$ model based
on the kinetic energy driven SC mechanism. Our result shows that
$t'$ plays an important role in enhancing the SC transition
temperature of the $t$-$J$ model. Within the $t$-$t'$-$J$ model,
we show that the SC-state of cuprate superconductors is the
conventional BCS like, so that the basic BCS formalism is still
valid in quantitatively reproducing the doping dependence of the
effective SC gap parameter and SC transition temperature, and
electron spectral function, although the pairing mechanism is
driven by the kinetic energy by exchanging dressed spin
excitations, and other exotic magnetic properties are beyond the
BCS theory.

Superconductivity in cuprates emerges when charge carriers, holes
or electrons, are doped into Mott insulators
\cite{kastner,tokura}. Both hole-doped and electron-doped cuprate
superconductors have the layered structure of the square lattice
of the CuO$_{2}$ plane separated by insulating layers
\cite{kastner,tokura}. In particular, the symmetry of the SC order
parameter is common in both case \cite{tsuei,tsuei6}, manifesting
that two systems have similar underlying SC mechanism. On the
other hand, the strong electron correlation is common for both
hole-doped and electron-doped cuprates, then it is possible that
superconductivity in electron-doped cuprates is also driven by the
kinetic energy as in hole-doped case. Within the $t$-$t'$-$J$
model, we \cite{feng5} have discussed this issue, and found that
in analogy to the phase diagram of the hole-doped case,
superconductivity appears over a narrow range of the electron
doping concentration in the electron-doped side, and the maximum
achievable SC transition temperature in the optimal doping in the
electron-doped case is much lower than that of the hole-doped side
due to the electron-hole asymmetry.

\acknowledgments The author would like to thank Dr. Huaiming Guo,
Professor Y.J. Wang, and Professor H.H. Wen for the helpful
discussions. This work was supported by the National Natural
Science Foundation of China under Grant Nos. 10125415 and
90403005.

\end{document}